\documentstyle[prd,aps]{revtex}

\newcommand{\Lag}{{\cal L}}
\newcommand{\argxst}{[x;\sta]}
\newcommand{\radij}{{\bf r}}
\newcommand{\vp}{{\bf p}}
\newcommand{\vk}{{\bf k}}

\newcommand{\tnic}{t_0}
\newcommand{\RR}{{\rm R}}

\newcommand{\RRRR}{\RR^{1,3}}
\newcommand{\scf}{\varphi}
\newcommand{\vef}{a}
\newcommand{\fvf}{\psi}
\newcommand{\fvfa}{\bar\psi}
\newcommand{\fvfF}{\widetilde\psi}
\newcommand{\staF}{\widetilde\sta}
\newcommand{\scfF}{\widetilde\varphi}
\newcommand{\vefF}{\widetilde a}
\newcommand{\GF}{\widetilde\G}
\newcommand{\jF}{\widetilde j}
\newcommand{\scfw}{\varphi_w}
\newcommand{\scsf}{q_0}

\newcommand{\scs}{j}
\newcommand{\G}{G}
\newcommand{\GW}{\G_w}
\newcommand{\GFW}{\GF_w}
\newcommand{\sta}{\Psi}
\newcommand{\Scat}{{\bf S}}

\newcommand{\Q}{{\bf Q}}
\newcommand{\p}{{\bf p}}
\newcommand{\intp}{\int \!d^4p}
\newcommand{\intpp}{\int \!d^4p'}
\newcommand{\psq}{p^2}
\newcommand{\ksq}{k^2}
\newcommand{\ppsq}{p^{\prime 2}}

\newcommand{\cnsa}[3]{a^{#1#2}_{#3}}

\newcommand{\half}{{1\over 2}}
\newcommand{\ksl}{{/\kern-6pt k}}

\newcommand{\ldp}{\vbox{\ialign{\hfil$##$\hfil\crcr
   \scriptscriptstyle\leftrightarrow\crcr\noalign{\kern.5pt\nointerlineskip}
   \partial\crcr}}}
\newcommand{\asta}{\overline{\sta}}

\begin{document}
\draft

\title{Framework for a theory that underlies the standard model}

\author{Marijan Ribari\v c and Luka \v Su\v ster\v si\v
c\cite{email}}

\address{Jo\v zef Stefan Institute, p.p.3000, 1001 Ljubljana,
Slovenia}

\maketitle

\begin{abstract}We put forward the following, physically motivated
premise for constructing a theory that underlies the standard model
in four-dimensional space-time: The Euler-Lagrange equations of such
a theory formally resemble some equations of motion underlying
fluid-dynamics equations in the kinetic theory of gases. Following
this premise, we point out Lorentz-invariant Lagrangians whose
Euler-Lagrange equations contain a subsystem equivalent to the
Euler-Lagrange equations of the standard model with covariantly
regularized propagators. 
\end{abstract}


\section{Introduction}

The standard model, which provides an adequate description of all
quantum-mechanical experiments so far performed, is generally
considered to be only an effective field theory: a low-energy
approximation to an underlying theory (UT). \cite{Weinb} To obtain a
UT, within the last fifteen years or so considerable effort has been
put into various string theories. But it is still open whether this
``top-down'' approach leads to the observed low-energy physics. In
this paper we put forward a new framework for an opposite,
``bottom-up'' approach to construction of a UT in four-dimensional
space-time. We base it on the analogy of the kinetic theory of gases
and give basic assumptions in Sec.~II, the premise in Sec.~III, and a
transport-theoretic example in Sec.~IV. As a starting point, we note
that: 

(A) Propagators of the standard model must be regularized to obtain
physically meaningful results. \cite{Weinb,Cheng} Following Pauli,
\cite{Villa} we presume there is a UT whose propagators (i)~do not
need to be regularized, and (ii)~can be regarded as such
regularizations of standard-model propagators that reflect
high-energy physics. \cite{Isham} In Sec.~IVC we construct a possible
Lagrangian of such a UT.

(B) Ever since Einstein, Podolski and Rosen published their
gedanken-experiment some sixty years ago, physicists have been aware
that, if we go beyond a strictly operational description of quantum
phenomena, interpretations of certain results suggest the existence
of {\it faster-than-light effects} (FTLEs). \cite{Sudbe,cvek1} So 
we expect the mathematical formalism of a UT to exhibit some FTLEs.
However, special relativity poses two serious conceptual problems in
connection with FTLEs:
\begin{itemize}
\item[(i)] When the relation between two events suggests FTLEs,
different observers may not agree on what is the cause and what is
the consequence! Suppose we observe two spacelike-separated
measurements, say A and B, and we believe that the result of the
earlier one determines the result of the latter one by a FTLE, e.g.,
by an instantaneous change of the quantum-mechanical state. In our
frame of reference, let A precede B so that we believe B is
determined by A. However, there are inertial frames where B precedes
A, and there observers believe the opposite, that B determines A.

\item[(ii)] It is not clear how to model states that exhibit FTLEs
without predicting that the present can influence the past! Suppose
the relation between responses and their sources is covariant in the
sense that to Lorentz-transformed source corresponds
Lorentz-transformed response. So, if a part of the response due to
some source is faster than light, then the corresponding parts of
responses to sources that equal certain Lorentz transformations of
this source precede their causes. Which is a very strong objection to
FTLEs, since no physical phenomenon ever suggested the existence of
``effects'' that precede their causes.
\cite{cvek1}
\end{itemize}
Resolution of such problems is often seen as the key to a better
understanding of quantum phenomena. \cite{Sudbe} We see no way around
the first problem. \cite{Aharo} Regarding the second problem, we
point out in Sec.~IV such Euler-Lagrange equations where one can
avoid this problem without coming in conflict with special relativity. 

\section{Basic assumptions}

The shape of a UT is quite unknown. However, we assume that {\it its
Euler-Lagrange equations are local\/} \cite{cvek2} {\it and
covariant, and the free-field equations} (i.e., Euler-Lagrange
equations with all non-linear terms taken as external sources) {\it
admit classical solutions that have: (i)~properties that are
propagated not faster than light according to covariant, regularized
Green functions of basic field equations, (ii)~unbounded front
velocity, and (iii)~no ``effects'' that precede their causes.} We
believe therefore that an understanding of classical systems with
such equations of motion would be invaluable in searching for and
constructing a UT. 

To this end let us be more specific about the mathematical properties
we expect from the above classical, free-field solutions in a
particular, scalar system: 
\begin{itemize} 
\item[(A)] An external source is described by a real function
$\scs(x)$ of the space-time variable $x = (ct, \radij) \in \RRRR$.
The set of possible sources is invariant under the inhomogeneous
Lorentz transformations $x \to \Lambda x + a$, i.e., if a particular
source $\scs(x)$ belongs to this set, then so does any ``moving''
source $\scs(\Lambda x + a)$.
\item[(B)] The total response of the system to the external source
$\scs(x)$ is described by the state function determined by some
local, linear, covariant, Euler-Lagrange equations of motion and
subsidiary conditions. A certain part of this total response,
described by the real function $\scf(x)$ of $x \in \RRRR$, is such
that: (i)~There is a Green function $\G(x)$ such that $\scf(x) = \G
\ast \scs$, where $\ast$ denotes the convolution with respect to $x$.
(ii)~The relation between $\scf(x)$ and $\scs(x)$ is covariant [if
$\scs(x)$ is the source of $\scf(x)$, then a ``moving'' source
$\scs(\Lambda x + a)$ causes $\scf(\Lambda x + a)$]. (iii)~Responses
$\scf(x)$ do not precede their sources $j(x)$. Thus, $\G(x)$ is
covariant in the sense that $\G(\Lambda x) = \G(x)$, and $\G(x) = 0$
if $t < 0$, and also if $c^2 t^2 < |\radij|^2$. \cite{mi002}
Consequently, (i)~all responses $\scf(x)$ exhibit Einstein's
causality, \cite{cvek1} and (ii)~the propagator
\begin{equation}
	\GF(k) \equiv \int d^4 x\, e^{-ik\cdot x} \G(x)
	\label{FourG}
\end{equation}
where $k \in \RRRR$ and $k\cdot x = \vk\cdot \radij - k^0 ct$, is
covariant, i.e., $\GF(\Lambda k) = \GF(k)$.
\item[(C)] This propagator $\GF(k)$ (i)~can be adequately
approximated by the propagator
\begin{mathletters}
\begin{equation}
	\GFW(k) \equiv ( \ksq )^{-1} 
	\label{Wprop}
\end{equation}
up to some extremely large value of $|\ksq |$, and (ii)~is regular in
the sense that
\begin{equation}
	\GF(k) = O\bigl( (\ksq )^{-n} \bigr)
	\label{regul}
\end{equation}
\end{mathletters}
as $\ksq \to \infty$, with constant $n > 2$. \cite{Cheng,Villa}

\item[(D)] The total response to any source $\scs(x)$ exhibits some
FTLEs, but no ``effects'' that precede their causes. So state
functions of the system in question depend causally on their sources;
but, as pointed out in Sec.~I, this dependence cannot be covariant!
Thus the subsidiary conditions that together with covariant equations
of motion determine the state functions cannot be covariant.
\item[(E)] All inertial frames are equivalent: The properties (A) to
(D), and the relations between state functions, responses $\scf(x)$,
and their sources $\scs(x)$ do not depend on the inertial frame of
the observer. Sources $\scs(x)$ and responses $\scf(x)$ are
relativistic: functions $\scs(x)$ that represent the same source in
different inertial frames are related by Lorentz transformations for
scalar fields; and the same goes for the corresponding functions
$\scf(x)$, in agreement with assumption~(B). As certain Lorentz
transformations of any state function exhibit effects that precede
their causes, total responses are not relativistic---the relation
between state functions in different inertial frames is open.
\cite{Aharo}
\end{itemize}

By (C) above, the Fourier transform $\GFW(k)$ of the wave-equation
Green function $\GW(x)$ adequately approximates $\GF(k)$ up to
extremely large values of $|\ksq |$. So, the wave response $\scfw(x)
\equiv \GW \ast \scs$ is a very good approximation to the response
$\scf(x)$ when the source $\scs(x)$ is varying slowly enough, both
spatially and temporaly. In such a case, the wave equation $( c^{-2}
\partial^2 /\partial t^2 - \vec\nabla \cdot \vec\nabla) \scfw(x) =
\scs(x)$ is a good approximation to the unknown, covariant equation
of motion for response $\scf(x)$. So we can say that the unknown
equations of motion for the state function of the system in question
underly the wave equation in the sense that their causal solutions,
though exhibiting FTLEs, propagate certain effects by a covariant,
regular propagator $\GF(k)$ that can be approximated by the wave
propagator $\GFW(k)$ up to extremely high values of $|\ksq |$.

\section{Premise}

Above properties (A) to (E), however, do not even suggest whether
such a classical system has a state function of only four continuous
variables ($ct$ and $\radij$), and certainly give no indication about
the nature of its equations of motion. So we went looking for
classical physical systems that behave similarly in order to
construct by analogy such equations of motion that underly the basic
free-field equations in the above sense. We found such systems in the
kinetic theory of gases. \cite{mi002}

Take a non-relativistic gas, for example. Its macroscopic state is
described by macroscopic variables such as kinetic energy and
density, slow changes of which propagate with a finite speed of
sound, approximately according to some fluid-dynamics
partial-differential equations. But its microscopic state is affected
almost immediately everywhere by a localized source since the
velocity of gas particles is not bounded in the non-relativistic
theory. Only a finite number of local averages of the microscopic
state are regarded as macroscopically observable, i.e., the
macroscopic variables, which can also be defined independently with
no reference to the microscopic state. The remaining properties of
the microscopic state (i.e., infinitely many, macroscopically
directly unobservable degrees of freedom), describe processes that
manifest themselves (i)~in fluctuations of macroscopic variables, and
(ii)~in the fact that fluid-dynamics equations are only
asymptotically valid approximations for smoothly and slowly changing
macroscopic variables.

For a rare gas of identical pointlike particles, the fluid-dynamics
equations can be extended to model somewhat faster changes of
macroscopic variables by introducing additional fields of spacetime
variable, which have no direct significance within the framework of
fluid dynamics, though they can be interpreted as local averages of
the microscopic state, see, e.g., the Grad method of moments.
\cite{Willi} But eventually these equations of motion cannot be
improved this way  any more, and one must resort to a more detailed
description by the one-particle distribution, a function of time,
position, and velocity, evolving according to the
integro-differential Boltzmann equation. \cite{Willi} So in the case
of a rare gas, there are a characteristic length and time interval
where a completely new physics appears with three additional
independent variables: physics essentialy different from the
macroscopic physics described by fluid-dynamics equations.

For various theoretical reasons, many theorists believe that the
framework of present quantum field theories may not be appropriate
for a theory of quantum phenomena valid for all energies. It was
Feynman who first suggested in Ref.~\onlinecite{Feynm} that the basic
partial-differential equations of theoretical physics might be
actually describing macroscopic motion of some infinitesimal entities
he called X-ons. In addition, already Heisenberg \cite{Heise} and
Bjorken and Drell \cite{Bjork} expected that there is a
characteristic energy (and length) beyond which quantum dynamics will
be essentially different from the one described by the canonical
formalism; so we expect the Euler-Lagrange equations of a UT to be
very different from those of the standard model. All of which,
together with the behaviour of a nonrelativistic gas, leads us to put
forward the following premise for a ``bottom-up'' approach to
fundamental interactions: {\it The Euler-Lagrange equations of a UT
formally resemble some of equations of motion underlying
fluid-dynamics equations in the kinetic theory of gases.} We believe
that this physically motivated premise will help us (i)~to construct
a UT whose propagators do not need to be regularized, and (ii)~to
model quantum-mechanical FTLEs. 

\section{Transport-theoretic example}

\subsection{Equation of motion}

Following the above premise, we now consider a class of systems with
properties (A) to (E), defined by covariant, linear,
integro-differential equations of motion with a non-covariant
causality condition. On the analogy with the linearized Boltzmann
equation let us provisionally regard these equations of motion as
modeling transport of some infinitesimal entities, X-ons, with
arbitrary four-momenta, whose macroscopic motion evolves almost
according to the wave equation.

We describe the state of X-ons in a given inertial frame by a real
state function $\sta(x,p)$ of the space-time variable $x \in \RRRR$
and of the four-momentum variable $p = (p^0, \p) \in \RRRR$.
\cite{cvek8} As the equation of motion for $\sta(x,p)$ we take the
local, linear, transport equation
\begin{equation}
	p{\cdot} \nabla \sta = \Scat\sta + \Q \,,
        \label{gtreq}
\end{equation}
where: (i)~$p{\cdot} \nabla = p^0 c^{-1} \partial/\partial t + \p
\cdot \vec\nabla$ is the covariant, substantial time derivative. Thus
the equation (\ref{gtreq}) with $\Scat = 0$ and $\Q = 0$ is an analog
of Newton's first law and describes free streaming of X-ons.

\noindent (ii)~The scattering operator $\Scat$ describes the
scattering of X-ons by the host medium---the vacuum. In the case
considered,
\begin{eqnarray}
	\Scat\sta &\equiv& f_0(\psq)\intpp f_0(\ppsq) 
		\sta(x,p') 
	\nonumber\\
	&&\qquad + f_1(\psq) p\cdot \intpp f_1(\ppsq) p' 
		\sta(x,p') - t(\psq) \sta(x,p) \,,
	\label{S-11l}
\end{eqnarray} 
where $f_0(\psq)$, $f_1(\psq)$ and $t(\psq)$ are real functions of
$\psq \in \RR$; and the integral
\begin{equation}
	\intp \,F(p) \equiv -i \lim_{r\to \infty} 
		\int_{-ir}^{ir} \kern-.5em dp^0 
		\int_{\psq \le r^2} F(p) \, d^3\!\p    		
	\label{inpdf}
\end{equation}
for functions $F(p)$, $p \in \RRRR$. \cite{cvek3}

\noindent (iii)~The source of all X-ons described by $\sta(x,p)$ is
given by
\begin{equation}
	\Q(x,p) \equiv \scsf f_0(\psq) \scs(x)  \,,
	\label{Qpdep}
\end{equation}
with $\scsf$ being a real parameter. As we do not permit effects that
precede their causes we assume the causality condition: if $\Q(x,p) =
0$ for all $t \le \tnic$, the corresponding state function
\begin{equation}
	\sta(x,p) = 0 \qquad\hbox{for all}\qquad t \le \tnic \,.
	\label{stcon}
\end{equation}

The equation of motion (\ref{gtreq}) is covariant with respect to
inhomogeneous Lorentz transformations
\begin{equation}
	x \to \Lambda x + a\,, \quad
	p \to \Lambda p\,, \quad
	\sta(x,p) \to \sta(\Lambda x + a, \Lambda p) \,,\quad
	\Q(x,p) \to \Q(\Lambda x + a, \Lambda p) \,.
	\label{trLot}
\end{equation}
However, like the Boltzmann equation, equation (\ref{gtreq}) is not
invariant under time reversal: state function $\sta(x,p)$ displays an
arrow of time in the sense that the time-reversed $\sta(x,p)$ is not
a solution to (\ref{gtreq}) with time-reversed source $\Q(x,p)$. In
contrast to the Einstein causality condition, condition (\ref{stcon})
is not covariant. As a consequence, the relation between solutions
$\sta(x,p)$ to equation (\ref{gtreq}) and their sources $\Q(x,p)$
need not be covariant.

\subsection{Properties of the state function $\lowercase{\sta(x,p)}$}

The total responses $\sta(x,p)$ of the system in question to sources
$\scs(x)$ are such that certain local averages, the macroscopic
variables
\begin{eqnarray}
	\scf[x;\sta] &\equiv& \int d^4p\, f_0(\psq) \sta(x,p) \,,
	\label{scsig} \\
	\vef[x;\sta] &\equiv& \int d^4p\, f_1(\psq) \sta(x,p) p\,,
	\nonumber
\end{eqnarray}
covariantly depend on sources $\scs(x)$ and exhibit Einstein's
causality despite the non-covariant causality condition
(\ref{stcon}). To infer this we proceed as in Ref.~\onlinecite{mi002}
to compute the Fourier transforms $\scfF[k;\sta]$ and $\vefF[k;\sta]$
of $\scf[x;\sta]$ and $\vef[x;\sta]$, and conclude that
\begin{mathletters}
\label{scmacvar}
\begin{eqnarray}
	\scfF[k;\sta] &=& \GF_0(k) \jF(k) \,,  \\
	\vefF[k;\sta] &=& ik \GF_1(k) \jF(k) \,,
\end{eqnarray}
where 
\begin{eqnarray}
	\GF_0(k) &=& q_0 D^{-1} (1 - I_3 - D ) \,, \\
	\GF_1(k) &=& q_0 D^{-1} I_2 \,,
\end{eqnarray}
with
\begin{eqnarray}
	D &\equiv& (1 - I_1)(1 - I_3) + \ksq I_2^2 \,, \\
	I_1(\ksq) &\equiv& (2\pi^2/\ksq) \int_0^\infty f_0^2(y)
		t(y) [ \sqrt{ 1 + \ksq y/t^2(y)} - 1 ] dy \,, \\
	I_2(\ksq) &\equiv& (\pi/\ksq)^2 \int_0^\infty f_0(y) f_1(y)
		t^2(y) [ \sqrt{ 1 + \ksq y/t^2(y)} - 1 ]^2 dy \,, \\
	I_3(\ksq) &\equiv& (\pi/\ksq)^2 \int_0^\infty f_1^2(y)
		t^3(y) [ \sqrt{ 1 + \ksq y/t^2(y)} - 1 ]^2 dy \,.
\end{eqnarray}
\end{mathletters}
As $\GF_0(k)$ and $\GF_1(k)$ are covariant, the corresponding
retarded Green functions $\G_0(x)$ and $\G_1(x)$ are covariant, and
$\scf[x;\sta] = \G_0\ast \scs$ and $\vef[x;\sta] = \nabla \G_1\ast
\scs$ satisfy Einstein's causality condition. \cite{mi002}

When $t^2(\psq)/\psq$ and its inverse are bounded for all $\psq \ge
0$, propagator $\GF_0(k)$ has the following properties:
\begin{itemize}
\item[(A)] If $\cnsa 114 \ne 4$, $\cnsa 002 = 1$ and $(4 -
\cnsa 114 )( 4\scsf - \cnsa 004 ) = ( \cnsa 014 )^2$, where
$\cnsa mnr \equiv \pi^2 \int_0^\infty f_m(y) f_n(y) [t(y)]^{m + n +
1} |\sqrt y / t(y)|^r dy $, then $\GF_0(k) = 1/\ksq + O((\ksq)^0)$ 
as $\ksq \to 0$.
\item[(B)] $\GF_0(k) = O((\ksq)^{-n})$ as $\ksq \to \infty$, where
$n = 1$ if $\cnsa 001 = 0$, $n=3/2$ if also $2\cnsa 000 = - (\cnsa
012)^2$, $n=2$ if also $ \cnsa 00{-1} = -4\cnsa 012 \cnsa 011
$, $n=5/2$ if also $ \cnsa 112 \cnsa 000 = 2 (\cnsa 011 )^2 + 2
\cnsa 012 \cnsa 010 $, and $n = 3$ if also $\cnsa 00{-3} = 8 \cnsa 
012 \cnsa 01{-1} + 32 \cnsa 011 \cnsa 010 - 16 \cnsa 000 \cnsa 111 -
4 \cnsa 00{-1} \cnsa 112$. 
\end{itemize}
These results enable us to explicitly show that within the presented
transport-theoretical framework there are covariant propagators
$\GF_0(k)$ regularizing the wave propagator $\GFW(k)$. Namely, when
$\sqrt{\psq}/t(\psq)$ has only two values for $\psq \ge 0$, say
$\tau_1$ and $-\tau_2$, we can explicitly calculate the corresponding
propagator $\GF_0(k)$ as a rational function of $\sqrt{ 1 +
\tau_j^2 \ksq}$, $j = 1$, $2$, whose six parameters are determined by
integrals of $f_0(\psq)$ and  $f_1(\psq)$. For $\tau_2 > \tau_1 > 0$,
there are infinitely many $f_0$ and $f_1$ such that for some real
$\scsf$ the corresponding $\GF_0(k)$ has the required properties:
(i)~it satisfies conditions (A) and (B) with $n=3$, (ii)~$\GF_0(k)$
is a decreasing function of $\ksq > 0$, (iii)~the difference
$|\GF_0(k) - \GFW(k)|$ is a bounded function of real $\ksq$, e.g.,
for $\tau_2/\tau_1 =2$, and (iv)~for any $\mu_0$ this difference can
be made arbitrarily small for all $|\ksq | < \mu_0$ by taking
$\tau_1$ and $\tau_2$ sufficiently small.

By (\ref{gtreq}), (\ref{S-11l}), (\ref{inpdf}), (\ref{stcon})
and (\ref{scsig}), when $\scs(x) = 0$ if $t \le \tnic$, we can
express the state function $\sta(x,p)$ for $p^0 \ne 0$ in terms of
the source $\scs(x)$ and fields $\scf\argxst$ and $\vef[x;\sta]$:
\cite{mi002}
\begin{equation}
	\sta(x,p) = \Theta(t - \tnic) \int_0^{c(t-\tnic)/p^0} 
		e^{- t(\psq) y} q(x - y p, p) \,dy 	
	\label{rw1sl}	
\end{equation}
with
\begin{equation}
	q(x,p) \equiv \{ \scf\argxst + \scsf \scs(x) \} f_0(\psq) 
		+  p\cdot \vef[x;\sta] f_1(\psq) 
	\label{QQdef}
\end{equation}
and $\Theta(t < 0) \equiv 0$ and $\Theta(t \ge 0) \equiv 1$. By
(\ref{scsig})--(\ref{QQdef}), the source $\scs(x)$ at $x = (ct_1,
\radij_1)$: (i)~does not affect $\sta(x,p)$ at $x = (ct_2,
\radij_2)$, $t_2 <t_1$, i.e., the system considered is causal; and
(ii)~affects $\sta(x,p)$ at $x = (ct_2, \radij_2)$, $t_2 > t_1$, for
some values of four-momentum $p$ no matter how small is the time
interval $t_2 - t_1$ and/or how large is the distance $|\radij_2 -
\radij_1|$, \cite{cvek4} i.e., the physical system considered
displays {\it everywhere arbitrary fast effects\/}: the front 
velocity of its state function $\sta(x,p)$ is not bounded! Thus the
dependence of $\sta(x,p)$ on $\scs(x)$ is not covariant in contrast
with the dependence of its properties $\scf[x;\sta]$ and
$\vef[x;\sta]$: the {\it covariance} (\ref{trLot}) of the equation of
motion (\ref{gtreq}) {\it is partly broken} by the non-covariant
causality condition (\ref{stcon}).

Regarding the connection between descriptions of X-ons in different
inertial frames, we assume that (i)~there is no prefered inertial
frame, (ii)~the source $\scs(x)$ is a scalar relativistic field, and,
(iii)~the independent variable $p$ transforms as a four-momentum. In
particular, when considering X-ons from an inertial frame whose
space-time coordinates $x' = \Lambda x + a$, their four-momenta $p' =
\Lambda p$, and their state function $\sta'(x',p')$ is uniquely
determined by (i)~the equations of motion
(\ref{gtreq})--(\ref{Qpdep}) with $\nabla \to \nabla'$, $p \to p'$ 
and $\scs(x) \to \scs'(x') = j(\Lambda^{-1}x' - \Lambda^{-1} a)$,
and (ii)~the non-covariant causality condition (\ref{stcon}). The
preceding results imply that $\G_0'(x) = \G_0(x)$ and $\G_1'(x) =
\G_1(x)$ so that $\scf'[x';\sta'] = \scf[x;\sta]$ and
$\vef'[x';\sta'] = \Lambda \vef[x;\sta]$; so these two local
averages of the state function are relativistic scalar and vector
fields. The state function itself is not relativistic; $\sta'(x',p')$
is related to $\scf'[x';\sta']$, $\vef'[x';\sta']$ and $\scs'(x')$
through the non-covariant relation (\ref{rw1sl}).

Above results show that one can construct integro-differential
equations of motion that can be regarded as underlying the wave
equation in the sense specified at the end of Sec.~II. In the same
manner one can construct also integro-differential equations that can
be regarded as underlying other basic, differential free-field
equations (see Appendices A and B).

\subsection{Lagrangian in accordance with the premise}

The equations of motion (\ref{gtreq})--(\ref{Qpdep}) equal the
Euler-Lagrange equations of the local, Lorentz-invariant Lagrangian
\begin{mathletters}
\label{Lagra}
\begin{equation}
	\Lag_0 = \Lag_{0 tr} + \Lag_{0 s} \,,
\end{equation}
with
\begin{eqnarray}
	\Lag_{0 tr}(\sta) &\equiv& (2\scsf)^{-1} \int\! d^4 p
		\, \sta(x,-p) [ p \cdot \nabla \sta - \Scat\sta ] \,, \\
	\Lag_{0 s}  &\equiv& - \int\! d^4 p \, \sta(x,-p) 
		f_0(\psq) \scs(x) \,.
\end{eqnarray}
\end{mathletters}
To construct a possible Lagrangian for a UT, we may proceed as
follows:
\begin{itemize}
\item[(i)] We take the Euler-Lagrange equations of the standard model
and express them in terms of spin-0, spin-$1\over 2$ and spin-1
propagators. 
\item[(ii)] We replace these propagators with propagators analogous
to $\GF_0(k)$ with properties~(B) to obtain relations such as
(\ref{scmacvar}) with spin-$0$, spin-$1\over 2$ and spin-$1$ sources.
\item[(iii)] Combining Lagrangians that are related to the obtained
relations as (\ref{Lagra}) is related to (\ref{scmacvar}), we can then
construct a possible transport-theoretic Lagrangian for a UT as
specified in Secs.~II and III. Its local and covariant Euler-Lagrange
equations comprise transport equations such as (\ref{gtreq}) with
scalar, spinor, and vector sources.
\end{itemize}
For QED, an example of such a construction is given in Appendix~C.
The question remains, however, which of the infinity of such
transport-theoretic Lagrangians are physically relevant for
constructing a UT. We considered quantum field theories defined by 
Feynman path integrals of such transport-theoretic Lagrangians in
Ref.~\onlinecite{cvek6}. It may be that only two functions of $x$ and
$p$ are needed for modeling of quantum phenomena: a four-vector one,
containing all integer-spin fields of fundamental forces, and a
chiral-bispinor one, containing all fields of fundamental matter
particles, see Ref.~\onlinecite{cvek6}.

The Euler-Lagrange equations of a Lagrangian constructed as specified
above contain a subsystem of equations equivalent to the
Euler-Lagrange equations of the standard model with some covariantly
regularized propagators. This subsystem determines the dynamics of
all fields of the standard model in the classical approximation so
that they exhibit no FTLEs, though solutions to the whole set of
covariant transport-theoretic Euler-Lagrange equations do exhibit
FTLEs. As in the classical approximation the temporal dependence of
the fields of the standard model describes the temporal dependence of
its quantum states, FTLEs are absent there. How to use
transport-theoretic FTLEs to explain FTLEs implied by certain quantum
phenomena is open. Such an explanation would not require, as sometime
suggested, \cite{mnogo} that we abandon the traditional belief that
the basic equations of motion are covariant, and all inertial frames
are equivalent. 

\section{Concluding remarks}

In this paper we have put forward a new framework for constructing a
theory that may underly the standard model. It requires that the
Euler-Lagrange equations of this theory: (i)~are local and covariant,
(ii)~have propagators that need not be regularized, (iii)~describe
some faster-than-light effects, and (iv)~formally resemble equations
of motion of some theory that underlies fluid dynamics in the kinetic
theory of gases. Motivations for and details of this framework are
given in Secs.~I--III.

To show that the proposed framework for modeling fundamental
interactions is feasible, we have pointed out in Sec.~IVC how one can
construct Euler-Lagrange equations such that: (i)~they are
integro-differential equations defined in eight-dimensional $\RRRR
\times \RRRR$ on the analogy with the Boltzmann transport equation,
(ii)~their causal solutions display FTLEs, and (iii)~certain local
averages of these solutions are propagated not faster than light by
the Euler-Lagrange equations of the standard model whose propagators
are covariantly regularized. Given transport-theoretic Euler-Lagrange
equations define for the first time {\it such a physically motivated
class of classical models that (i)~are not invariant under time
reversal, (ii)~have covariant, regular propagators, (iii)~model
certain FTLEs without predicting that present can influence the past,
and (iv)~are not in conflict with special relativity.}

Whether it makes physical sense to interprete such
transport-theoretic Euler-Lagrange equations as describing the
macroscopic movement of some Feynman X-ons is an open question. It
took almost thirty years since the formulation of the kinetic theory
of gases by Maxwell and Boltzmann until the basic idea of molecules
was accepted as a physical reality due to Perrin's experimental work
that verified Einsteins's and Smoluckowski's analysis of Brownian
motion. So it would be of great interest if one could identify some
phenomenon characteristic of X-ons.

\section{Acknowledgements}

We are grateful to Matja\v z Polj\v sak and Igor Sega for
many useful suggestions.

\appendix
\section{Scalar boson propagator}

Propagator for a massive scalar field is $\G_{KG} \equiv (\ksq +
m^2)^{-1}$, $m \ge 0$, and equals the propagator for a spin-$1$
massive boson in the Feynman gauge. Propagator $\G_0(k)$ defined by
(\ref{scmacvar}a) can be made regular and approximate $\G_{KG}(k)$ as
accurately as desired for all $k$ up to some extremely large value of
$|\ksq|$ by choosing $q_0$, $f_0(y)$, $f_1(y)$ and $t(y)$ so that
(i)~conditions (B) in Sec.~IVB are satisfied, (ii)~at $y = -m^2$, 
\begin{mathletters}
\label{sccon}
\begin{eqnarray}
	[1 - I_1(y)][1 - I_3(y)] &=& y I_2^2(y) \,, \\
	q_0 [ 1 - I_3(y) ] &=& d \{ [1 - I_1(y)][1 - I_3(y)] +
		y I_2^2(y) \} /dy \,,
\end{eqnarray}
\end{mathletters}
and (iii)~$|t(y)|$ is sufficiently large.

\section{Spin $\half$ fermion propagator}

Let the Fourier transforms of the chiral bispinor field and source
$\fvfF(k)$ and $\fvfF_s(k)$, $k \in \RRRR$, be related by
spin-$\half$ fermion propagator
\begin{equation}
	\GF_{\half}(k) \equiv {m - i\ksl\over m^2 + k^2}\,,
	\qquad m \ge 0\,,
	\label{ferpr}
\end{equation}
i.e., let $\fvfF = \GF_{\half} \fvfF_s$.

We will consider a system whose state is described by the
bispinor-valued function $\sta_{\half}(x,p)$ of $x, p \in \RRRR$,
that is a solution to the Euler-Lagrange equations of the Lagrangian
\begin{mathletters}
\label{ferdf}
\begin{eqnarray}
	\Lag_{\half} &\equiv& \Lag_{\half tr} + \Lag_{\half s} \,, \\
	\Lag_{\half tr}(\sta_{\half}) &\equiv& q_{\half}^{-1}
		  \intp \, \asta_{\half}(x,-p) [ p^\mu \ldp_\mu 
		  + t(\psq) ] \sta_{\half}(x,p)
	\nonumber \\ & & \qquad {}-s q_{\half}^{-1} \intp \intpp \Bigl [
	f_0(\ppsq) f_1(\psq) p_\mu \asta_{\half} (x,-p) \gamma^\mu 
		\sta_{\half}(x,p') + {\rm c.c.} \Bigr ] \,,
	\label{ferLatr} \\
	\Lag_{\half s}(\sta_{\half}) &\equiv& -\intp \Bigl [
		f_0(\psq) \asta_{\half} (x,-p) \fvf_{s}(x) 
		+ {\rm c.c.} \Bigr ] \,, \label{ferLas}
\end{eqnarray}
\end{mathletters}
where: $2 a \ldp_\mu b \equiv a (\partial_\mu b) - (\partial_\mu a)
b$; $\asta_{\half} \equiv \sta_{\half}^\dagger i\gamma^0\,$;
$\gamma^\mu$ are the Dirac matrices; $t(\psq)$, $f_0(\psq)$, and
$f_1(p^2)$ are real-valued functions of $p^2 \in \RR$; and $s$ and
$q_{\half}$ are real parameters. Lagrangian $\Lag_{\half}$ is real
and changes sign under charge conjugation $\sta_{\half}(x,p) \to
\eta_c \gamma^2 \sta^*_{\half} (x,p)$ and $\fvf_{s}(x) \to \eta_c
\gamma^2 \fvf^*_{s}(x)$, $|\eta_c|^2 = 1$. It transforms as a scalar
field of $x$ under: (i)~Lorentz transformations, (ii)~spatial
inversion $\sta_{\half}(ct, \radij, p^0, \vp) \to \eta_p \gamma^0
\sta_{\half} (ct, -\radij, p^0, -\vp)$ and $\fvf_{s}(ct, \radij) \to
\eta_p \gamma^0 \fvf_{s}(ct, -\radij)$, $|\eta_p|^2 = 1$, and
(iii)~time reversal $\sta_{\half}(ct, \radij, p^0, \vp) \to \eta_t
\gamma^1 \gamma^3 \sta^*_{\half} (-ct, \radij, -p^0, \vp)$ and
$\fvf_{s}(ct,\radij) \to \eta_t \gamma^1 \gamma^3 \fvf^*_{s}(-ct,
\radij)$, $|\eta_t|^2 = 1$. 

We take Fourier transforms $\staF_{\half}(k,p)$ of the solution to the
Euler-Lagrange equations of $\Lag_{\half}$ that is subject to a
causality condition such as (\ref{stcon}) to define the chiral
bispinor field of $k \in \RRRR$:
\begin{equation}
	\fvfF_m(k) \equiv \intp \, f_0(\psq) \staF_{\half}(k,p) \,.
	\label{fvmdf}
\end{equation}
The relation between $\fvfF_m(k)$ and its source $\fvfF_s(k)$,
\begin{mathletters}
\label{spiprop}
\begin{equation}
	\fvfF_m(k) = \GF_{\half m}(k) \fvfF_s(k)
	\label{macpr}
\end{equation}
where
\begin{eqnarray}
	\GF_{\half m}(k) &=& N(\ksq) { M(\ksq) - i\ksl \over
		M^2(\ksq) + \ksq } \,,
	\label{fvfsp} \\
	N(\ksq) &\equiv& q_{\half} I_1(\ksq)/ 2s I_2(\ksq) \,,
	\label{fvfNdf} \\
	M(\ksq) &\equiv& \{ 1 - s^2 [ I_1(\ksq) I_3(\ksq) 
		+ \ksq I_2^2(\ksq)]\} \Big/ 2s I_2(\ksq) \,,
	\label{fvfMdf}
\end{eqnarray}
with $I_1(\ksq)$ and $I_2(\ksq)$ given by (\ref{scmacvar}f) and
(\ref{scmacvar}g), and
\begin{equation}
	I_3(\ksq) \equiv (2\pi^2/\ksq) \int_0^\infty y f_1^2(y)
		t(y) [ \sqrt{1+\ksq y/t^2(y)} - 1] dy \,,
	\label{defi3} 
\end{equation}
\end{mathletters}
is invariant under the charge conjugation transformation. For certain
$f_0(y)$, $f_1(y)$, $t(y)$, $q_{\half}$ and $s$, we can take
$\G_{\half m}(k)$ as a regularization of spin $\half$ propagator.
Namely, (i)~as $k^2 \to \infty$,
\begin{equation}
	\GF_{\half m}(k) = O( (k^2)^{-n} )
	\label{ferreg}
\end{equation}
with $n=1/2$ if $a_1 = 0$, $n=1$ if also $a_0= 0$, $n= 3/2$ if also
$a_{-1} = 0$, and $n = 5/2$ if also $a_{-3} = 0$, where $a_r \equiv
\int_0^\infty f_0^2(y) t(y) | \sqrt{y}/t(y) |^r dy$, (ii)~the difference
$| \GF_{\half m}(k) - \GF_{\half}(k) |$ can be made arbitrarily small
for all $|k^2| \le \mu_o^2$ for any $\mu_o^2 > 0$ provided $|t(y)|$
is suffficiently large and $q_{\half}$ and $s$ are such that $N(y) =
1 + d M^2(y)/dy$ and $M(y) = m$ at $y = -m^2$, and (iii)~$\GF_{\half 
m}(k)$ is bounded for all real $k^2 \ne -m^2$.

The propagator $\GF_{\half m}(k)$ is through multiplication by
$N(\ksq)$ regularized fermion propagator $\GF_{\half}(k)$ with
energy-dependent mass $M(k^2)$, which does not equal zero even when
the mass $m$ of the low-energy approximation $\GF_{\half}(k)$ equals
zero. So within the transport-theoretic framework presented the
neutrino mass may equal zero in the low-energy, quantum
field-theoretic approximation, though it is definitely not equal to
zero for higher energies! 

\section{Transport-theoretic regularization of propagators in
quantum electrodynamics}

Lagrangian of quantum electrodynamics in Feynman gauge reads
\begin{equation}
	\Lag_{QED} = -\half (\partial_\mu A_\nu)^2 - \fvfa (
		\gamma^\mu \ldp_\mu + ie \gamma^\mu A_\mu
		+ m) \fvf \,.
\end{equation}
Its Euler-Lagrange equations are the wave equation $\partial_\nu
\partial^\nu A^\mu = -i e \fvfa \gamma^\mu \fvf$ and the Dirac
equation $(\gamma^\mu \partial_\mu + m)\fvf = - ie\gamma^\mu A_\mu
\fvf$. 

Transport-theoretic, Lorentz-invariant Lagrangian
\begin{equation}
	\Lag_{ttQED} \equiv \sum_{\mu = 0}^3 \eta_{\mu\mu} \Lag_{0 tr}
		(\sta_\mu) + \Lag_{\half tr}(\sta_{\half}) 
		- ie \fvfa_m \gamma^\mu \scf[x;\sta_\mu] \fvf_m
\end{equation}
where $\sta_\mu(x,p)$ are components of a four-vector, has
Euler-Lagrange equations that contain in $k$-space the following
subsystems:
\begin{eqnarray}
	\scfF[k;\sta_\mu] &=& -ie \GF_0(k) \bar{\fvfF_m}(k) \gamma^\mu
		\ast \fvfF_m (k) \,, 		\\
	\fvfF_m(k) &=& - ie \G_{\half m}(k) \gamma^\mu \scfF[k;\sta_\mu]
		\ast \fvfF_m(k) \,.		
\end{eqnarray}
If we replace $\GF_0(k)$ with $\GF_w(k)$ and $\GF_{\half m}(k)$ with
$\GF_{\half}(k)$, these two subsystems become equivalent to the above
wave and Dirac equations. Under conditions given in Sec.~IVC and
Appendix~B, the transport-theoretic Lagrangian $\Lag_{ttQED}$ yields
Euler-Lagrange equations correspoding to those of $\Lag_{QED}$ with
regularized propagators.


\begin{references}
\bibitem{cvek1}A change in the state of a physical system that occurs
at the space-time point $(ct_2, \radij_2)$ and is attributed to a
source at $(ct_1, \radij_1)$ is refered to as: an effect (response)
if $t_2 \ge t_1$; a faster-than-light effect (FTLE) if $|\radij_2 -
\radij_1| > c(t_2 - t_1)$ and $t_2 \ge t_1$; an ``effect'' that
precedes its cause if $t_2 < t_1$; an effect that satisfies
Einstein's causality condition if $|\radij_2 - \radij_1| \le c(t_2 -
t_1)$.
\bibitem{cvek2}We label an equation of motion as local if it relates
only the values of the state function and of finitely many of its
time and space derivatives at the same space-time point.
\bibitem{cvek4}We can regard $c |\vp/p^0|$ as the speed of an $X$-on
with four-momentum $p$. If we take as the domain of the independent
variable $p$ the invariant subset $\{ p : p^0 > 0, |\vp/p^0| < 1 \}$
of $\RRRR$, then no $X$-on is faster than light, the relation
(\ref{rw1sl}) is covariant, the state function $\sta(x,p)$ defined by
(\ref{rw1sl}) exhibits Einstein's causality, and there are no FTLEs.
On the other hand, if $\Scat = 0$ and $\Q(x,p) = 0$ if $|\vp/p^0| \le
1$, then all X-ons are faster than light and all properties of
$\sta(x,p)$ propagate faster than light.
\end{references}
\end{document}